\documentclass[journal, onecolumn]{IEEEtran}
\IEEEoverridecommandlockouts
\usepackage{cite}
\usepackage{amsmath,amssymb,amsfonts}
\usepackage{algorithmic}
\usepackage{algorithm}
\usepackage{graphicx}
\usepackage{textcomp}
\usepackage{xcolor}
\usepackage{subfigure}
\def\BibTeX{{\rm B\kern-.05em{\sc i\kern-.025em b}\kern-.08em
		T\kern-.1667em\lower.7ex\hbox{E}\kern-.125emX}}
	
\begin{document}

\title{Toward Multiple Integrated  Sensing and Communication Base Station Systems: Collaborative Precoding Design with Power Constraint\\
	\thanks{This work was supported in part by the National Key Research and Development Program under Grant 2020YFA0711302. 
	Correspondence authors: Zhiqing Wei and Zhiyong Feng.}
}
\author{
Wangjun~Jiang,
Zhiqing~Wei and
Zhiyong~Feng
\\Key Laboratory of Universal Wireless Communications, Ministry of Education
\\Beijing University of Posts and Telecommunications, Beijing, China
\\Email: \{jiangwangjun, weizhiqing, fengzy\}@bupt.edu.cn
}

\maketitle

\begin{abstract}

The collaborative sensing of multiple Integrated sensing and communication (ISAC)
base stations is one of the important technologies to achieve intelligent transportation. 
Interference elimination between ISAC base stations is the prerequisite for realizing collaborative sensing.
In this paper, we focus on the mutual interference elimination problem in collaborative sensing of multiple ISAC base stations that can communicate and radar sense simultaneously by transmitting ISAC signals. We establish a mutual interference model of multiple ISAC base stations, which consists of communication and radar sensing related interference. Moreover, we propose a joint optimization algorithm (JOA) to solve the collaborative precoding problem with total power constraint (TPC) and per-antenna power constraint (PPC). The optimal precoding design can be obtained by using JOA to set appropriate tradeoff coefficient between sensing and communication performance.
The proposed collaborative precoding design algorithm is evaluated by considering sensing and communication performance via numerical results. 
The complexity of JOA for collaborative precoding under TPC and PPC is also compared and simulated in this paper. 

\end{abstract}


%
\IEEEpeerreviewmaketitle

\section{Introduction}
%
%
%
%

Integrated sensing and communication (ISAC) base stations are gradually becoming one of the important devices for intelligent transportation \cite{[PMC]}, which can communicate and radar sense simultaneously by transmitting ISAC signals. The idea of using multiple ISAC base stations for collaborative sensing was proposed by IMT 2030 working group \cite{[IMT_2030]}. 
An important prerequisite for collaborative sensing of multiple ISAC base stations is to eliminate the mutual interference between ISAC base stations. 
Compared to traditional base stations, the mutual interference in ISAC base station systems is more complex, which contains three parts:
1) downlink (DL) communication related interference includes multi-user interference and the inter-base station interference,
2) radar sensing related interference includes multipath echo interference of base station itself and inter-base station echo interference,
Collaborative precoding design is one of the key technologies to solve the interference problem. 
3) the interference between sensing and communication.

For multi-user interference, M. Schubert {\it {et al.}} first proposed a solution to the precoding design problem of eliminating multi-user interference \cite{[SDP_37]}. 
For inter-base station interference, the spectral efficiency of forced zero precoding and dirty paper coding in base station collaboration was investigated in \cite{[CoMP_20]}. However, the computational complexity of the participating base stations in \cite{[CoMP_20]} is high. Recognizing this fact, the distributed precoding design method was presented for collaborative base stations, which make full use of the computational power of each base station \cite{[CoMP_22]}.
For radar sensing related interference, the main solutions to the precoding design problem of eliminating the interference include chirp shifting, phase scrambling and frequency hopping \cite{[Radar_1]}. A conventional adaptive interference phase-cancellation algorithm was proposed by using a large aperture auxiliary antenna array \cite{[Radar_29]}. S. Zhang proposed a time-frequency filtering method based on the short-time Fourier transform to effectively filter out the interference signal with linear/curvilinear distribution in the time-frequency domain \cite{[Radar_71]}.

Compared with traditional base stations, the mutual interference in ISAC base station systems is more complex, including not only the communication and sensing related interference, but also the interference between sensing and communication. 
J. Andrew {\it {et al.}} proposed a procedure of communication and sensing in the time division duplex (TDD) mode, which can effectively avoid the interference between sensing and communication, i.e., the interference between the uplink (UL) communication signal and ISAC echo signal \cite{[TDD]}. 
However, the interference between ISAC base stations is not considered in the above study, which will be further studied in this paper. 

\subsection{Main Contributions of Our Work}

To analyze the interference between ISAC base stations, we establish a mutual interference model between multiple ISAC base stations, and put forward a joint optimization algorithm (JOA) for collaborative precoding design under total power constraint (TPC) and per-antenna power constraint (PPC). 
{
The existing cooperative precoding technology is mainly used to solve the interference problem in the communication system, but this paper focuses on the interference problem of ISAC base stations. The interference between ISAC base stations is more complicated, not only between communications, but also between sensing. In this paper, the ISAC mutual interference model is proposed to analyze the complex interference between ISAC base stations, and JOA algorithm is proposed to jointly optimize the precoding of ISAC base station signals to improve communication and sensing performance. The simulation results show that the performance of sensing and communication need to be weighed in the actual collaborative precoding process. The tradeoff coefficient between sensing performance and communication performance is set according to the actual scene requirements, and the JOA algorithm proposed in this paper is used to achieve collaborative precoding, so as to obtain the expected sensing performance and communication performance.}
For clarity, we list the contributions of this paper as follows:

1. A mutual interference model including DL communication related interference and radar sensing related interference is established in this paper. DL communication interference includes multi-user interference and inter-base station interference. Radar sensing related interference includes multipath echo interference and inter-base station echo interference. 

2. JOA for collaborative precoding design under TPC and PPC is put forward in this paper, which transforms the collaborative precoding problem into a convex problem by omitting the rank constraint.

The remaining parts of this paper are organized as follows.
Section \ref{sec:system_model} describes the system model of multiple ISAC base stations and the mutual interference model. The collaborative precoding design problem is formulated in Section \ref{sec:PF}. Section \ref{sec:JOA} introduces JOA for collaborative precoding design under TPC and PPC. The improvement of sensing and communication performance based on proposed collaborative precoding design is analyzed and simulated in section \ref{sec:Simulation}. Section \ref{sec:Conclusion} concludes the paper.

\section{System Model of ISAC base stations}\label{sec:system_model}

\subsection{Multiple ISAC Base Station System}\label{sec:system_model_1}

\begin{figure}[ht]
	\includegraphics[scale=0.23]{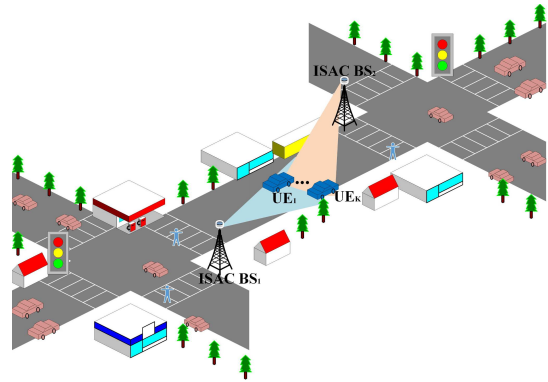}
	\centering
	\caption{Multiple ISAC base stations system.}
	\label{fig:ISAC_system_model}
\end{figure}

The scenario studied in this paper is ISAC base station sensing road condition information.
As Fig. \ref{fig:ISAC_system_model} shows, ISAC $\rm{BS}_1$ and ISAC $\rm{BS}_2$ simultaneously send ISAC signals ${\bf X}_1$ and ${\bf X}_2$ to the target user for DL communication and sensing. Assuming that ISAC base stations with $N_t$ antennas are providing communication services to $K$ single-antenna users while detecting the target user.

\subsection{ISAC Mutual Interference Model}\label{sec:system_model_2}

\begin{figure}[ht]
	\includegraphics[scale=0.18]{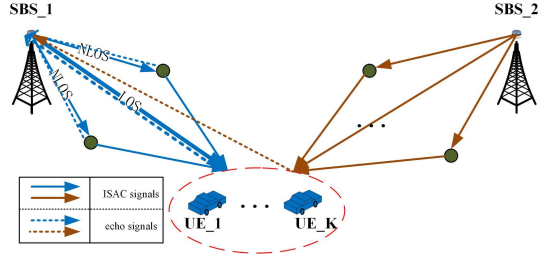}
	\centering
	\caption{ISAC mutual interference model.}
	\label{fig:ISAC_matual_model}
\end{figure}

This subsection will introduce the received signal models of users and ISAC base stations respectively, and then establish the ISAC mutual interference model.

 \subsubsection{Received Signal Model of users}\label{sec:system_model_2_1}
 
 The received signal of users contains three parts, ${{\bf H}{_1}}{{\bf X}{_1}}$: the desired signal from ISAC $\rm{BS}_1$,
 ${{\bf H}{_2}}{{\bf X}{_2}}$: the interference signal from ISAC $\rm{BS}_2$ and ${{{\bf Z}_C}}$: the communication noise. 
 Therefore, the received signal of users can be expressed as
 \begin{equation}\label{equ:Received_Signal_Model_1}
 {{{\bf Y}_{UE}}}  = {{\bf H}{_1}}{{\bf X}{_1}} + {{\bf H}{_2}}{{\bf X}{_2}} + {{{\bf Z}_C}},
 \end{equation}
 where
 \begin{equation} \label{equ:Received_Signal_Model_2}
 \begin{aligned}
 {{\bf H}{_1}} &= [{\bf h}_{1,1},{\bf h}_{1,2},…,{\bf h}_{1,K}]^T, \in {{\mathcal C}^{K\times N_t}} \\
 {{\bf H}{_2}} &= [{\bf h}_{2,1},{\bf h}_{2,2},…,{\bf h}_{2,K}]^T, \in {{\mathcal C}^{K\times N_t}}
 \end{aligned}
 \end{equation}
 are the DL communication channel of ISAC $\rm{BS}_1$ and ISAC $\rm{BS}_2$, respectively.
 
 After considering multi-user interference, the received signal of the $i$th user can be expressed as
 \begin{equation}
 \begin{aligned}
{{{\bf Y}_{UE_i}}} = {\bf h}_{1,i}{{\bf x}_{1,i}} +  \overbrace{{\bf h}_{2,i}{\bf X}_2 + {\bf Z_{C_i}} +\sum_{j=1,j \ne i}^{K} { {\bf h}_{1,j}}{{\bf x}_{1,j}}}^{\rm interference}.
 \end{aligned}
 \label{equ:Received_Signal_Model_7}
 \end{equation}
 
 After considering inter-base station interference, the Signal to Interference plus Noise Ratio (SINR) of the received signal of the $i$th user can be derived as
 \begin{equation}\label{equ:Received_Signal_Model_8}
 \begin{aligned}
 {\gamma_{C_i}} = & \frac{ |{\bf h}_{1,i} {{\bf x}_{1,i}}|^2 }{ |{ {\bf h}_{2,i}}{\bf X}_2|^2 + {{\bf \sigma}^2_{C_i}} +\sum_{j=1,j \ne i}^{K} |{\bf h}_{1,j}{{\bf x}_{1,j}}|^2 } \\
 = & \frac{ {\rm tr}({\bf h}_{1,i}^H {\bf h}_{1,i} {{\bf T}_{1,i}})  }   { {\rm tr}({\bf h}_{2,i}^H {\bf h}_{2,i} {\bf X}_2 {\bf X}_2^H ) + \sum_{j=1,j \ne i}^{K} {\rm tr}({{\bf h}_{1,j}^H {\bf h}_{1,j}} {{\bf T}_{1,j}})    +  {{\bf \sigma}^2_{C_i}} }
 \end{aligned},
 \end{equation}
 where ${\rm tr}(\bf X)$ denotes the trace of $\bf X$,  ${\bf \sigma}^2_{C_i}$ is the power of the communication noise of the $i$th user, ${{\bf T}_{1,i}} = {{\bf x}_{1,i}} {{\bf x}_{1,i}^H} \succeq 0$ and $ rank ({{\bf T}_{1,i}}) = 1$.
 
 \subsubsection{Received Signal Model of ISAC $\rm{BS}_1$}\label{sec:system_model_2_2}
 
 The received signal of the ISAC $\rm{BS}_1$ contains three parts, ${{\bf G}{_1}}{{\bf X}{_1}}$: the desired echo signal from ISAC $\rm{BS}_1$,
 ${{\bf G}{_2}}{{\bf X}{_2}}$: the interference echo signal from ISAC $\rm{BS}_2$ and ${{{\bf Z}_R}}$: the radar sensing noise and ground clutter. Therefore, the received signal of the ISAC $\rm{BS}_1$ can be expressed as
 \begin{equation}
 \begin{aligned}
 {{{\bf Y}_{BS}}} =& {{\bf G}{_1}}{{\bf X}{_1}} + {{\bf G}{_2}}{{\bf X}{_2}} + {{{\bf Z}_R}} \\
  = &  {{\bf g} {_{1,0}}}{{\bf X}{_1}} + \overbrace{{{\bf G}{_2}}{{\bf X}{_2}} + {{\bf Z}_R} + \sum_{l=1}^{L_{p}-1} {{{\bf g}{_{1,l}}}{{\bf X}{_1}}}}^{\rm interference},
 \end{aligned}
 \label{equ:Received_Signal_Model_2_2}
 \end{equation}
 where ${{\bf G}{_1}}$ and ${{\bf G}{_2}}$ are the radar channels of ISAC $\rm BS_1$ and ISAC $\rm BS_2$, ${{\bf g}{_{1,0}}}$ is the Line of Sight (LOS) path of ${{\bf G}{_1}}$, ${{\bf g}{_{1,l}}}$ is the other paths of ${{\bf G}{_1}}$. The number of signal-dependent interference sources is $L_p$ \cite{[SC_37]}.
 Then SINR of the received signal of ISAC $\rm BS_1$ can be derived as
 \begin{equation}\label{equ:Received_Signal_Model_2_3}
 \begin{aligned}
 {\gamma_{R}} &= \frac{ |{{\bf g}{_{1,0}}}{{\bf X}_1}|^2  }{ |{{\bf G}{_2}}{{\bf X}{_2}}|^2 + {{\bf \sigma}_R^2} + \sum_{l=1}^{L_{p}-1} |{{{\bf g}{_{1,l}}}{{\bf X}{_1}}}|^2 } \\
 &= \frac { {\rm tr}({{\bf g}{_{1,0}^H}}{{\bf g}{_{1,0}}} {{\bf X}_1} {{\bf X}_1^H} )}   { {\rm tr}({{\bf G}{_2^H}} {{\bf G}{_2}}{{\bf X}{_2}} {{\bf X}{_2^H}} ) + \sum_{l=1}^{L_{p}-1}  { {\rm tr}({{\bf g}{_{1,l}^H}}  {{\bf g}{_{1,l}}} {{\bf X}_1} {{\bf X}_1^H} )} +  {\bf \sigma}_R^2 }
 \end{aligned}.
 \end{equation}
 where ${{\bf \sigma}_R^2}$ is the power of the radar channel noise.

\section{Problem Formulation}\label{sec:PF}

In this section, we will formulate the collaborative precoding problem with power constraints. Both TPC and PPC are considered in this section.
The collaborative precoding design problem for ISAC base stations is how to meet the SINR constraints and power constraints with minimum power.
The SINR constraints can be expressed as 
\begin{equation}
\begin{aligned}
\frac { {\rm tr}({{\bf g}{_{1,0}^H}} {{\bf g}{_{1,0}}} {{\bf X}_1} {{\bf X}_1^H})  } { {\rm tr}( {{\bf g}{_{2,0}^H}} {{\bf g}{_{2,0}}}{{\bf X}{_2}} {{\bf X}{_2^H}})  + \sum_{l=1}^{L_{p}-1} {\rm tr}( {{\bf g}{_{1,l}^H}}{{\bf g}{_{1,l}}} {{\bf X}_1} {{\bf X}_1^H} )  + {\bf \sigma}_R^2 } \ge {\zeta }_{R}
\end{aligned},
\label{equ:SINR_PF_1}
\end{equation}
\begin{equation}
\begin{aligned}
\frac{ {\rm tr}({\bf h}_{1,i}^H {\bf h}_{1,i} {{\bf T}_{1,i}})  }   { {\rm tr}({\bf h}_{2,i}^H {\bf h}_{2,i} {\bf X}_2 {\bf X}_2^H ) + \sum_{j=1,j \ne i}^{K} {\rm tr}({ {\bf h}_{1,j}^H {\bf h}_{1,j}} {{\bf T}_{1,j}})    +  {{\bf \sigma}^2_{C_i}} } \ge {\zeta }_{C,i}
\end{aligned},
\label{equ:SINR_PF_2}
\end{equation}
where ${\zeta }_{R}$ is the SINR threshold of ISAC $\rm BS_1$, { ${\zeta }_{C,i}$} is the SINR threshold of the $i$th user. 
TPC can be expressed as
\begin{equation}
\begin{aligned}
& \quad \frac{1}{L} ||{{\bf X}_1}||_F^2 \le {P_t} \\
& \quad \frac{1}{L} ||{{\bf X}_2}||_F^2 \le {P_t}
\end{aligned},
\label{equ:TPC_PF}
\end{equation}
PPC can be expressed as
\begin{equation}
\begin{aligned}
& \quad \frac{1}{L}  {diag({{\bf X}_1}{{\bf X}_1}^H)} \le \frac{P_t}{N_t} {\bf I}_{N_t} \\
& \quad \frac{1}{L}  {diag({{\bf X}_2}{{\bf X}_2}^H)} \le \frac{P_t}{N_t} {\bf I}_{N_t}
\end{aligned},
\label{equ:PPC_PF}
\end{equation}
where ${\bf I}_{N_t}$ denotes the ${N_t} \times {N_t}$ identity matrix.

{
The purpose of collaborative precoding is to minimum the interference between ISAC base stations.  
Therefore, the problem of collaborative precoding design is how to minimum the interference between ISAC base stations with TPC or PPC being the constraint. Since the interference can be represented by ${\gamma_{R}}$ and ${\gamma_{C_i}}$, the problem can be transformed as how to maximize ${\gamma_{R}}$ and ${\gamma_{C_i}}$ with TPC or PPC being the constraint. The purpose of the problem is to maximize ${\gamma_{R}}$ and ${\gamma_{C_i}}$.  According to \eqref{equ:SINR_PF_1} and \eqref{equ:SINR_PF_2}, we can turn \eqref{equ:SINR_PF_1} and \eqref{equ:SINR_PF_2} into SINR constraints by setting thresholds. Then, the problem of collaborative precoding design can be converted to how much power should ISAC base station transmit to meet SINR constraints, TPC and PPC. Therefore, the problem of collaborative precoding design can be expressed as
\begin{equation}
\begin{aligned}
{\mathcal{P}_{1}} : \min_{{{\bf X}_1}, {{\bf X}_2}} & \quad P_t  \\
s.t. 
& \quad {\gamma_{R}} \ge {\zeta }_{R} \\
& \quad {\gamma_{C,i}} \ge {\zeta }_{C,i} \\
& \quad {{\bf T}_{1,i}} \succeq 0, {{\bf T}_{1,i}} = {{\bf T}_{1,i}^H} \\
& \quad  rank({{\bf T}_{1,i}}) = 1, \forall i \\
& \quad  \eqref{equ:TPC_PF} \quad for \quad TPC   \\
& \quad \eqref{equ:PPC_PF}  \quad for \quad PPC  \\ 
\end{aligned}.
\label{equ:PF_1}
\end{equation}

}
\section{JOA for Collaborative Precoding Design}\label{sec:JOA}

\begin{algorithm}[htb]  
	\caption{Joint optimization algorithm (JOA)}  
	\label{alg:JOA_1}  
	\begin{algorithmic}  
		\REQUIRE
		$\bf Input$: $\bf H$, $P_t$, ${\zeta }_{C,i}$, ${\zeta }_{R}$ and $N = N_t = N_r$;\\
		\ENSURE ${\bf X}_1$ and ${\bf X}_2$;
		\IF {PPC}
		\STATE 1). Divide the diagonal constraint of TPC into \\ \quad N quadratic equality constraints.
		\ENDIF
		\STATE \quad 2). Solve the SDP problem ${\mathcal{P}_{1}}$ by omitting the rank-1 
		\STATE \quad \quad constraint;
		\STATE \quad 3). Obtain the approximated solution by eigenvalue 
		\STATE \quad \quad decomposition or Gaussian randomization.
	\end{algorithmic}  
\end{algorithm}

{
This section will introduce JOA for collaborative precoding design. Since the constraint of $rank({{\bf T}_{1,i}}) = 1$ is non-convex, problem ${\mathcal{P}_{1}}$ is can not be solved directly by using convex optimization. 
Nonetheless, ${\mathcal{P}_{1}}$ can be a standard SDP by omitting the constraint of $rank({{\bf T}_{1,i}}) = 1$, which can be solved by the classic semidefinite program (SDP) technique \cite{[SDP_37],[SDP_92]}. 
The approximated solution can be obtained by standard rank-1 approximation techniques, such as eigenvalue decomposition or Gaussian randomization \cite{[liu_seperation_38]}.
According to \cite{[SDP_37]}, the global optimum can be achieved by SDP technique, so the approximated solution obtained by standard rank-1 approximation techniques is also approximately the global optimal solution.

}
The proposed JOA can be summarized by the following Algorithm \ref{alg:JOA_1}.

\section{Simulation Results}\label{sec:Simulation}

In this section, we will first analyze the communication and sensing performance of the collaborative precoding design under TPC and PPC and then analyze the trade-off between the communication and sensing performance. In the end, we further compare the complexity of JOA for collaborative precoding design under TPC and PPC. 
Simulation parameters used in this section are shown in table \ref{Parameter:simulation} \cite{[Liu_parameters_1],[Strum_parameters_1]}.

\begin{table}[t]
	\caption{Simulation parameters adopted in this paper.}
	\label{Parameter:simulation}
	\begin{tabular}{l|l|l|l|l|l}
		\hline
		\hline
		Items & Value & Meaning of the parameter & Items & Value & Meaning of the parameter \\ \hline
		$f_c$ & $24$ GHz \cite{[Strum_parameters_1]} & Frequency of ISAC signals & $B$ & 100 MHz \cite{[Strum_parameters_1]} & System bandwidth\\ \hline
		$N=N_t=N_r$ & 10 \cite{[Liu_parameters_1]}& Number of antenna array elements & $K$ & 5 \cite{[Liu_parameters_1]}& Number of users\\ \hline
		$L$ & 50 \cite{[Liu_parameters_1]}& Length of data stream & $L_p$ & 3 & Number of interference sources \\ \hline
		$\theta_k$ & $10^o\sim20^o$ & Angle of users relative to BS & $P_t$ & $20$ dBm \cite{[Strum_parameters_1]}& Transmitting power of BS\\ \hline
		$\bf H$ & $\mathcal{C}^{K*N_t}$ & Communication channel & $\bf G$ & $\mathcal{C}^{K*N_t}$ & Radar channel \\ \hline
		${\bf \sigma}_R$ & -94 dBm & Thermal noise of radar channel & ${\bf \sigma}_C$ & -94 dBm & Noise of communication channel  \\ \hline
		${\zeta}_R$ & $0\sim30$ dB & Threshold of radar SINR & ${\zeta}_{C}$ & $0\sim30$ dB & Threshold of communication SINR \\ \hline
	\end{tabular}
\end{table}

\subsection{Communication and Sensing Performance}\label{sec:Simulation_1}

\begin{figure}[ht]
	\includegraphics[scale=0.4]{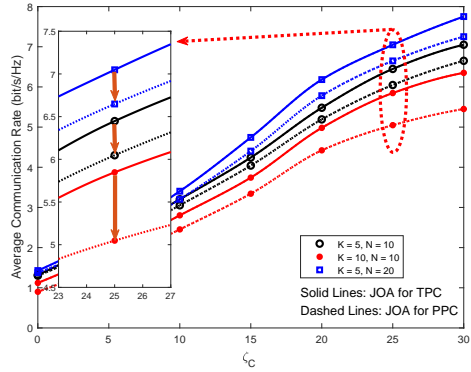}
	\centering
	\caption{Average communication rate for different ${\zeta}_C$ with ${\zeta}_R=10$ dB.}
	\label{fig:C_1}
\end{figure}

\begin{figure}[ht]
	\includegraphics[scale=0.4]{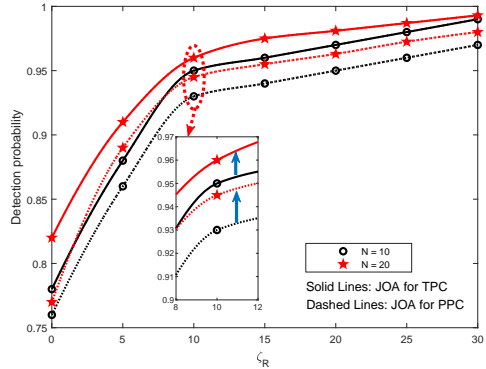}
	\centering
	\caption{Detection probability for different ${\zeta}_R$ with ${\zeta}_C=10$ dB.}
	\label{fig:PD_1}
\end{figure}

\begin{figure}[ht]
	\includegraphics[scale=0.4]{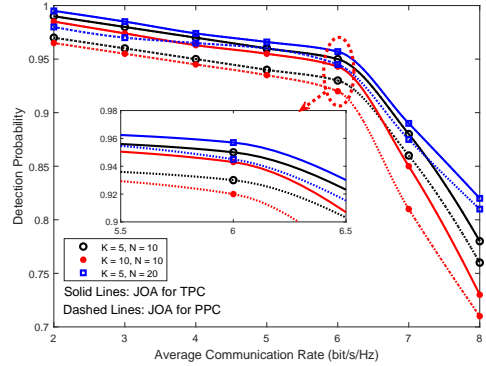}
	\centering
	\caption{Trade-off between communication rate and detection probability.}
	\label{fig:C_PD_1}
\end{figure}

{
\subsubsection{Average communication rate}\label{sec:Simulation_1_1}

Assuming that the communication SINR threshold of all users is the same, i.e ${\zeta}_{C} = {\zeta}_{C,i}, \forall i$. The average communication rate $\bar{C}$ can be defined as
\begin{equation}
\begin{aligned}
\bar{C} = \frac{\sum_{i=1}^{K}{log_2 (1+\gamma_{C_i})} }{K}
\end{aligned}.
\label{equ:Sim_1}
\end{equation}

As Fig. \ref{fig:C_1} shows, the collaborative precoding design with TPC can obtain a relatively higher average communication rate than that with PPC. Since PPC is more stringent than TPC, the collaborative precoding design with TPC has a higher degree of freedom and thus achieves a higher communication rate, which is consistent with the simulation results. As the number of users $K$ increases, the SINR constraint is more stringent, the average communication rate obtained by collaborative precoding decreases. Moreover, since PPC will also be more stringent with the increase of $K$, the decreasing trend of average communication rate obtained by collaborative precoding based on PPC is more obvious than that of TPC. As for the number of antenna array elements $N$, the average communication rate obtained by collaborative precoding will increase with the increase of $N$.

\subsubsection{Detection probability}\label{sec:Simulation_1_2}

For collaborative precoding design under TPC and PPC, the detection probability $P_d$ is used as the metric, where we consider the constant false alarm rate detection for point-like targets from $L_p-1$ path. The false alarm probability for ISAC base station is $P_f = 10^{-7}$. And we can obtain the detection probability $P_d$ based on \cite{[PD_9]} eq. (69). 
Fig. \ref{fig:PD_1} shows that the detection probability is increasing with increasing ${\zeta}_R$ based on ${\zeta}_C=10$ dB. Moreover, the detection probability will also increase with the increase of antenna array element number $N$.

\subsubsection{Trade-off between Communication Performance and Sensing Performance}\label{sec:Simulation_2}

As Fig. \ref{fig:C_PD_1} shows, there exists a trade-off between the communication rate and the radar detection performance. the communication rate decreases with the increase of the $P_d$. The trend is reflected in both JOA for TPC and PPC with different number of users $K$ and antenna array elements $N$. The optimal precoding design can be obtained by using JOA to set appropriate tradeoff coefficient between sensing and communication performance.
 
}

\subsection{Comparison of the  Complexity of JOA for TPC and PPC}\label{sec:Simulation_3}

\begin{figure}[ht]
	\includegraphics[scale=0.4]{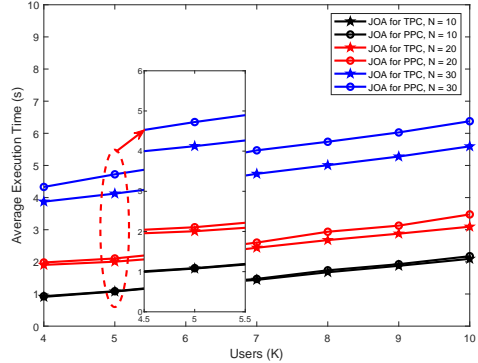}
	\centering
	\caption{Average execution time for different $K$ and $N$ with ${\zeta}_R=10$ dB and ${\zeta}_C=10$ dB.}
	\label{fig:Complexity_1}
\end{figure} 


{
	
In this subsection, we study the computational complexity of JOA for PPC and TPC, which is constrained by two aspects. One is the threshold constraint of SINR, the other is the constraint of the number of users $K$ and the number of antennas $N$. The former mainly affects the number of iterations of SDP optimization problem, while the latter mainly affects the complexity of matrix operation in each iteration process. Since there is no closed-form expression for the iterative complexity of SDP optimization, we use the average execution time to reflect it.
For the complexity of matrix operation in each iteration optimization process, we analyze it in terms of flops, which is defined as multiplication, division, or addition of two floating point numbers \cite{[Flop_43]}. 
Considering the flops of the complex addition and multiplication, both the complexity of JOA for TPC and PPC can be expressed as $\mathcal{O}(N^2K + NK^2)$, which means that the complexity of JOA for TPC and PPC have the same trend with respect to $N$ and $K$. 
According to section \ref{sec:JOA}, the complexity of JOA for PPC is higher than that for TPC because of the process of dividing the diagonal constraint of TPC into N quadratic equality constraints. To verify the conclusion, we compare the complexity performance of JOA for PPC and TPC by average execution time. It should be noted that due to the error of computer running time, the average execution time for each case is the average of multiple tests.
Fig. \ref{fig:Complexity_1} shows the average execution time for different number of users $K$ and antenna array elements $N$ with the threshold of radar SINR ${\zeta}_R=10$ dB and communication SINR ${\zeta}_C=10$ dB. We find that as the number of antenna array elements $N$ and users $K$ increases, the average execution time of solving the precoding design optimization under TPC and PPC is increasing.

}

\section{Conclusion}\label{sec:Conclusion}
{
ISAC base stations can communicate and radar sense simultaneously by transmitting ISAC signals.
In this paper, we focus on the mutual interference elimination problem in collaborative sensing of multiple ISAC base stations.
We establish a mutual interference model of multiple ISAC base stations, which consists of communication and radar sensing related interference. Moreover, we propose JOA to solve the collaborative precoding problem with TPC and PPC. The optimal precoding design can be obtained by using JOA to set appropriate tradeoff coefficient between sensing and communication performance.
Simulation results show that there is a trade-off between communication and sensing performance. Moreover, the comparison of the complexity of JOA for TPC and PPC is also simulated in this paper.
}


%

\bibliographystyle{IEEEtran} 
\bibliography{reference}	


\ifCLASSOPTIONcaptionsoff
  \newpage
\fi

\end{document}